\documentclass[aps,twocolumn,showpacs,superscriptaddress]{revtex4-1}
\usepackage{amsmath}
\usepackage{braket}
\usepackage{color}
\usepackage{amsfonts}
\usepackage{amssymb}
\usepackage{graphicx}
\usepackage[english]{babel}
\usepackage[colorlinks=true]{hyperref}
\usepackage{soul}

\begin{document}


\title{Inner-shell clock transition in atomic thulium with small BBR shift}


\author{A.\,Golovizin}
\author{E.\,Fedorova}
\affiliation{P.N.\,Lebedev Physical Institute, Leninsky prospekt 53, Moscow, 1119991, Russia}
\affiliation{Russian Quantum Center, Business-Center ``Ural'', 100A Novaya St., Skolkovo, Moscow 143025, Russia}
\author{D.\,Tregubov}
\affiliation{P.N.\,Lebedev Physical Institute, Leninsky prospekt 53, Moscow, 1119991, Russia}
\author{D.\,Sukachev}
\affiliation{P.N.\,Lebedev Physical Institute, Leninsky prospekt 53, Moscow, 1119991, Russia}
\affiliation{Russian Quantum Center, Business-Center ``Ural'', 100A Novaya St., Skolkovo, Moscow 143025, Russia}
\affiliation{Physics Department of Harvard Univeristy, 17 Oxford str., Cambridge, MA 02138, USA}
\author{K.\,Khabarova}
\author{V.\,Sorokin}
\affiliation{P.N.\,Lebedev Physical Institute, Leninsky prospekt 53, Moscow, 1119991, Russia}
\affiliation{Russian Quantum Center, Business-Center ``Ural'', 100A Novaya St., Skolkovo, Moscow 143025, Russia}
\author{N.\,Kolachevsky}\email{kolachevsky@lebedev.ru}
\affiliation{P.N.\,Lebedev Physical Institute, Leninsky prospekt 53, Moscow, 1119991, Russia}
\affiliation{Russian Quantum Center, Business-Center ``Ural'', 100A Novaya St., Skolkovo, Moscow 143025, Russia}


\date{\today}

\begin{abstract}
With direct polarizability measurements we demonstrated extremely low sensitivity of the inner-shell clock transition at $1.14\,\mu$m in Tm atoms to external dc electric fields and black-body radiation (BBR). 
We measured differential polarizabilities of clock levels in Tm at wavelengths of 810--860\,nm and at 1064\,nm and inferred the static scalar differential polarizability of the inner-shell clock transition of $-0.047(18)$ atomic units corresponding to only $2\times10^{-18}$ fractional frequency shift from BBR at the room temperature. 
This is a few orders of magnitude smaller compared to the BBR shift of the clock transitions in the neutral atoms (Sr, Yb, Hg) and competes with the least sensitive ion species (e.g. Al$^+$ or Lu$^+$). 
For the $1.14\,\mu$m clock transition, we experimentally determined the ``magic'' wavelength of $813.320(6)$\,nm, recorded the transition spectral linewidth of $10$\,Hz, and measured its absolute frequency of $262\,954\,938\,269\,213(30)$\,Hz.
\end{abstract}

\pacs{}

\maketitle

Unprecedented performance of the state-of-the-art optical atomic clocks \cite{Chen2017,Ushijima2015,Marti2018} together with the
increasing number of characterized clock transitions in different atomic and ionic species opened new frontiers in physics, namely,
sensitive tests of special relativity \cite{Delva2017,Shaniv2018}, search for  dark matter~\cite{roberts2017search,Wciso2018} and variations of fundamental constants \cite{Rosenband2008,Borkowski2018}, atom-based  quantum
simulations~\cite{Norcia259}, and some other as summarized in  reviews \cite{ Ludlow2015,Poli2014,safronova2018search}.
Moreover, development of robust, compact, and transportable optical clocks~\cite{Koller2017}  overcoming in accuracy the best
microwave frequency standards promises a new era in precision spectroscopy, navigation, and
geodesy~\cite{Grotti2018,takano2016geopotential}.

The fractional frequency instability of optical clocks reaching low $10^{-18}$ level  was successfully demonstrated for Sr
\cite{Nicholson2015} and Yb \cite{schioppo2017ultrastable} optical lattice clocks as well as for Yb$^+$~\cite{Huntemann2016} and
Al$^+$ ~\cite{Chou2010} single-ion optical clocks.
One of the current limitations is the impact of electric fields like  surrounding black body radiation (BBR), trapping  fields,
and collisions.  To push forward the current limit in frequency stability and accuracy, there is a continuous search for new species,
particularly
with the reduced sensitivity to external electric fields.
Among promising candidates are  single  Lu$^+$ ion~\cite{Arnold2018},
highly charged ions~\cite{nauta2017towards,Yu2018,Kozlov2018}, and $^{229}$Th with an isomeric nuclear
transition~\cite{Wense2017,thielking2018laser}.

Lanthanides with the submerged electronic $f$-shell possess natural suppression of the sensitivity to external electric fields for the inner-shell $f$-$f$ transitions because of strong shielding by the closed $5s^2$ and $6s^2$ shells.  
Already in 1984, E.\,Alexandrov and co-authors demonstrated unusually low spectral broadening of the 1.14\,$\mu$m inner-shell transition in atomic Tm under collisions with buffer He gas~\cite{aleksandrov1984eb}.
In 2004, strong shielding effect for Tm-He collisions was confirmed in ref.\,\cite{Hancox2004}.
Similar effect was also observed for some transition elements with nonzero orbital angular momentum~\cite{Hancox2005}.
For lanthanide ions doped in solids, the shielding effect reduces inhomogeneous broadening of the inner-shell $f$-$f$ transitions which allows, for example, ensemble-based solid-state quantum memory~\cite{probst2013} and integrated single-photon source in telecom wavelength range~\cite{Dibos2018}.
However, the possibility to use such transitions for optical clocks was studied only
theoretically~\cite{Kozlov2013,kozlov2014optical,Sukachev2016}.

In this Letter, we report on precision spectroscopy of the inner-shell clock transition $\ket{J=7/2,F=4,m_F=0} \rightarrow\ket{J=5/2,F=3,m_F=0}$ between the fine-structure components of the ground electronic state in atomic Tm at the wavelength of 1.14\,$\mu$m with the natural linewidth of $\gamma=1.2$\,Hz (here $J$ and $F$ stand for the electronic and total momentum quantum numbers, respectively, and $m_F$ is the magnetic quantum number).
Spectroscopy of the $\ket{m_F=0} \rightarrow \ket{m_F=0}$ clock transition leads to zero first order Zeeman and magnetic dipole-dipole interaction shifts~\cite{Sukachev2016}.
We experimentally determined magic wavelength of the optical lattice near 813\,nm, recorded Fourier-limited spectral linewidth of the clock transition of 10\,Hz and characterized its sensitivity to electric and magnetic fields.
Accurate  measurement of  differential dynamic polarizabilities of the clock levels in the near infrared spectral region (810--860\,nm and 1064\,nm) allowed to estimate the BBR frequency shift of the clock transition, which
turned out to be  a few orders of magnitude smaller compared to other characterized clock transitions in neutral atoms.

\begin{figure}[t!]
\includegraphics[width=1\linewidth]{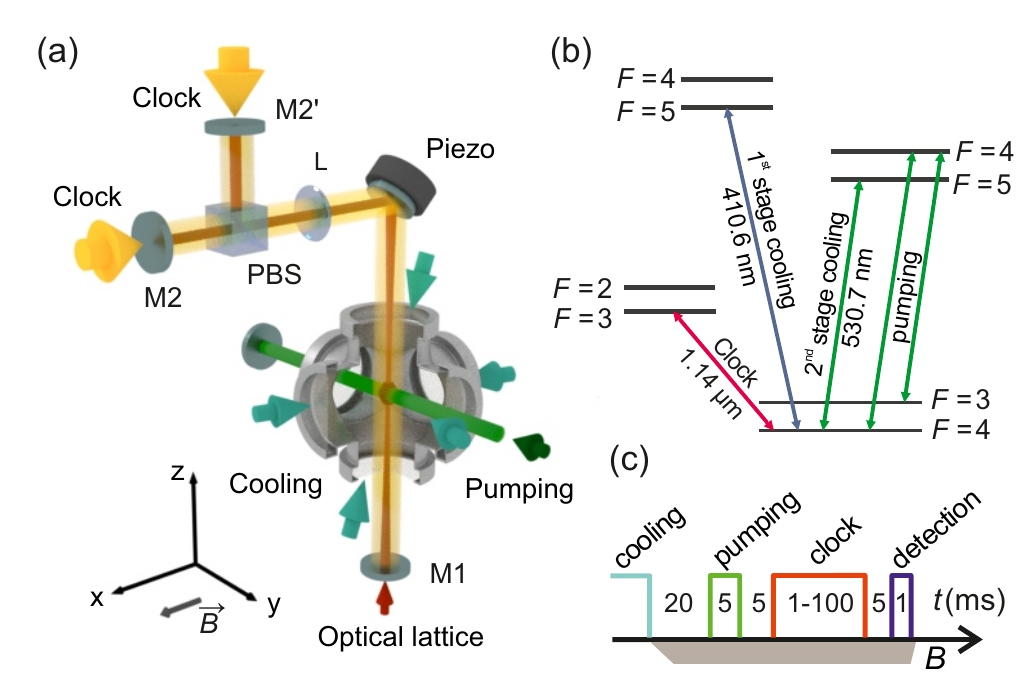}
\caption{ Experiment details.
a) Schematics of the setup: enhancement cavity forming the vertical optical lattice is built in a $\Gamma$-shape configuration
with an intra-cavity polarizing beam splitter (PBS). For different polarizations
two output couplers M2 and  M2$^\prime$ are used.
Clock laser radiation at 1.14\,$\mu$m is fed collinearly with the cavity axis through one of the cavity outcouplers with
corresponding
polarization;
pumping and repumping laser beams form standing waves along $y$-axis; initial position of the Tm magneto-optical trap is adjusted to
the cavity waist. M1:~cavity incoupler, L:~lens.
b) Relevant Tm level scheme.
c) Pulse sequence for spectroscopy of the clock transition.  Two stage laser cooling is followed by pumping to the $\ket{F=4,m_F=0}$
ground state using 530.7 nm radiation. The optical lattice in continuously on. After excitation of the clock transition one measures
the fluorescence of atoms remaining in the ground state by 1\,ms pulse at 410.6\,nm  (detection). The bias magnetic field
$B$ is denoted by a gray bar on the bottom.
}
\label{img:setup}
\end{figure}

Thulium atoms are laser cooled in a two-stage magneto-optical trap (see Fig.\,\ref{img:setup} and detailed description in ref.\,\cite{sukachev2014secondary}).
For precision spectroscopy, atoms at a temperature of $\sim10\,\mu$K are loaded in a vertical optical lattice formed inside an
enhancement cavity (finesse equals 20) with intra-cavity polarization filtering.
The cavity allows to build-up optical power from a 1\,W Ti:sapphire laser up to 6\,W in the wavelength range of 810--860\,nm.
The beam waist radius can be varied between 80\,$\mu$m and 160\,$\mu$m by controlling the distance between  cavity elements.

We prepare atoms in lower clock state $\ket{F=4,m_F=0}$ by driving simultaneously $\ket{F=4}\rightarrow\ket{F=4}$ (pump) and $\ket{F=3}\rightarrow\ket{F=4}$ (repump) transitions with a 5\,ms-long $\pi$-polarized laser pulses at 530.7\,nm   in the presence of a  bias magnetic field of $B_x\sim0.1$\,G.
After the optical pumping cycle,  $10^5$ atoms are trapped in the lattice with more than 80\% in the target state.
To interrogate the clock transition, we use radiation of a semiconductor $1.14\,\mu$m laser stabilized to a high-finesse Ultra Low Expansion glass (ULE) cavity, providing the
relative frequency instability of smaller than 10$^{-14}$ in 1-100\,s integration time \cite{Alnis2008}. The linear frequency drift of 29\,mHz/s is compensated  using an acousto-optical modulator. After compensation, the clock laser can be used as a stable  frequency reference for studying frequency shifts of the clock transition.

\subsection*{Results}

\begin{figure}[t!]
    \includegraphics[width=.9\linewidth]{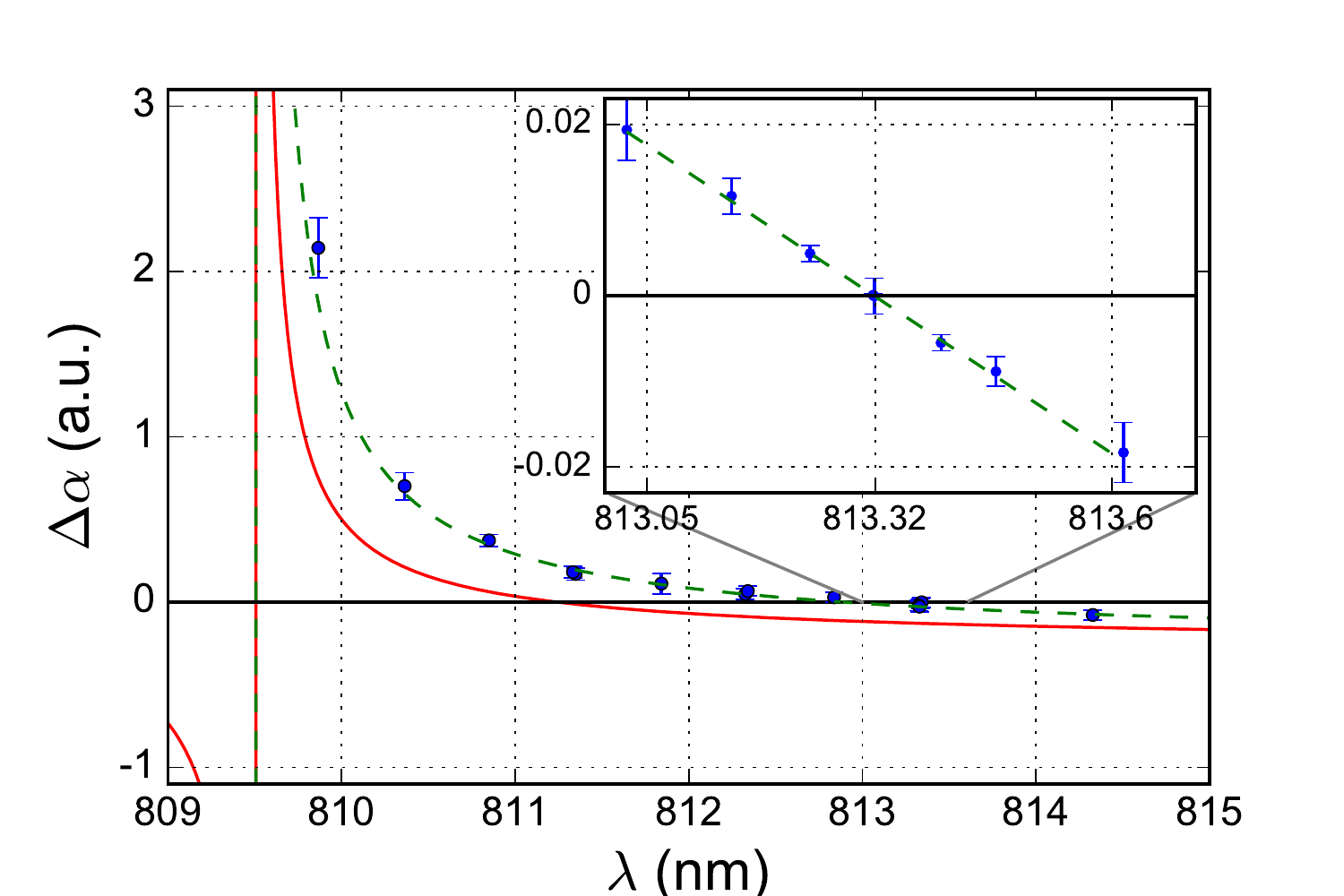}
\caption{Magic wavelength determination.
Calculated (Model\,1, red solid curve), measured (blue dots), and fitted (Model\,2, green dashed line)  differential dynamic polarizability $\Delta\alpha$
(in atomic units, a.u.) between the upper ($J=5/2$) and the lower ($J=7/2$) clock levels; theoretical models are described further in
the text and in Methods.
Inset:  zoom of the spectral region around the magic wavelength $\lambda_m=813.320(6)$\,nm.
}
\label{img:mwl}
\end{figure}

{\bf Magic wavelength determination.} The important step towards high resolution spectroscopy of the clock transition is determination of a magic wavelength of the optical lattice when  dynamic polarizabilitites of the clock levels become equal~\cite{Takamoto2003}.
We numerically calculated dynamic polarizabilities of the clock levels using time-dependent second order perturbation theory and transition data obtained using {\small COWAN} package \cite{cowan1981} (Model~1 described in Methods and in ref.\,\cite{Sukachev2016}), and  predicted existence of the magic wavelength at 811.2\,nm (for the collinear magnetic field $\vec{B}$ and the lattice field polarization $\vec{\epsilon}$)  near a narrow transition from the upper $J=5/2$ clock level at $809.5$\,nm.
Calculated differential polarizability is shown in Fig.\,\ref{img:mwl} with the red solid line. This wavelength region is readily accessible by Ti:sapphire or powerful semiconductor lasers.

Using an approach described in~ref.\,\cite{Barber2008}, we experimentally searched for the magic wavelength for the clock transition in the
spectral  region of 810--815\,nm.
The transition frequency shift  $\Delta\nu$ as a function of optical lattice power $P$ was measured at different lattice wavelengths.
The differential dynamic polarizability $\Delta\alpha$ between the clock levels was calculated using the expression $h\Delta\nu=-16 \,a_0^3 \Delta\alpha P /c \,w^2$. Here $h$ is the Plank constant, $c$ is the speed of light, $a_0$ is the Bohr radius, and $w=126.0(2.5)$\,$\mu$m is the lattice beam radius which was calculated from the enhancement cavity geometry.
The intra-cavity power $P$ was determined by calibrated photodiodes placed after the cavity outcouplers M2 and M2$^\prime$. 
The details of the beam waist determination and power measurements are given in Methods. 
Figure\,\ref{img:mwl} shows the spectral dependency of $\Delta\alpha$ in atomic units~(a.u.). 
The magic wavelength of $\lambda_m=813.320(6)$\,nm was determined by zero crossing of the linear fit in the inset of Fig.\,\ref{img:mwl}.

\begin{figure}
 \includegraphics[width=1\linewidth]{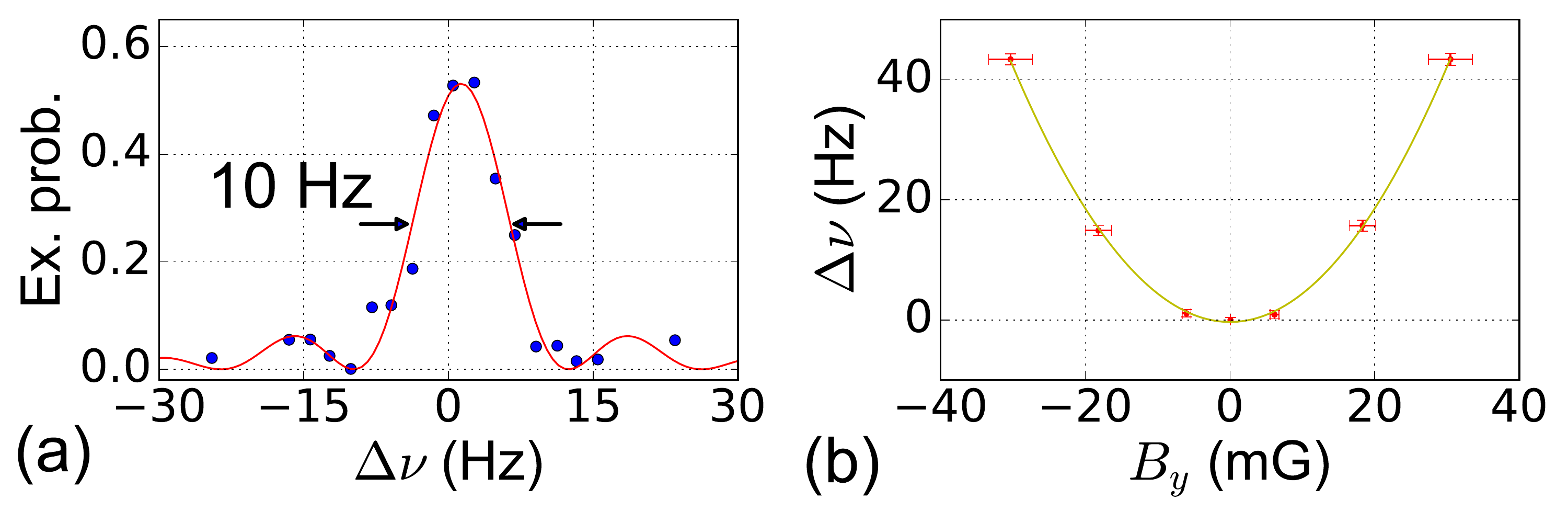}
\caption{Spectroscopy of the clock transition.
a)~Spectral line  shape of the clock transition in Tm. 
Every point is average of 6 measurements. 
The solid curve shows the fit calculated for a Fourier-limited $80$\,ms rectangular $\pi$-pulse.
b)~Clock transition frequency shift $\Delta\nu$ depending on  $B_y$ (dots) at $B_z=0$ and constant $B_x=225\,$mG; solid line is a parabolic fit.
The dependence on $B_z$ is similar.}
\label{img:zeeman}
\end{figure}

Trapping Tm atoms in the optical lattice at $\lambda_m$ drastically reduces inhomogeneous ac Stark broadening of the clock transition.
Exciting with a 80\,ms-long Rabi $\pi-$pulses of the clock laser we recorded a spectrum with 10\,Hz full width at the half maximum shown in Fig.\,\ref{img:zeeman}(a). 
Non-unity excitation at the line center comes from a non-perfect initial polarization of
atoms and  finite lifetime of the upper clock level ($\tau=112$\,ms).

Using Ti:sapphire frequency comb,  we determined the absolute frequency  of $\ket{J=7/2,F=4}\rightarrow\ket{J=5/2,F=3}$ clock transition in Tm of $262\,954\,938\,269\,213(30)$\,Hz.
The relative frequency uncertainty  of $1.1\times10^{-13}$  mainly comes from instability and calibration accuracy of a GLONASS-calibrated passive hydrogen maser used as a frequency reference for the comb.

{\bf Differential polarizability analysis.} In the second order approximation, the energy shift of an atomic level  $\ket{J,F,m_F}$ in an external oscillating electromagnetic field with the wavelength $\lambda$  equals  $-\alpha_{J,F,m_F}(\lambda)E^2/2$, where $E$ is the amplitude of the electric  field.
For linear field polarization, the dynamic polarizability $\alpha_{J,F,m_F}$ can be split in the scalar $\alpha_{J}^s$ and the tensor $\alpha_{J,F}^t$ parts as:
\begin{equation}
\label{eq:pol}
 \alpha_{J,F,m_F} = \alpha_{J}^s +\frac{3 \cos^2\Theta - 1}{2}\times\frac{3 m_F^2 - F (F+1)}{ F (2 F -1)} \alpha_{J,F}^t\,,
\end{equation}
where $\Theta$ is the  angle between the quantization axis (here the direction of the external magnetic field $\vec B$) and the
electric field polarization $\vec{\epsilon}$ of the optical lattice.
In our case, the differential polarizability of the two clock levels equals
\begin{equation}
\label{eq:pol3}
\Delta\alpha \equiv \alpha_{5/2,3,0}-\alpha_{7/2,4,0}= \Delta\alpha^s + \frac{3 \cos^2\Theta - 1}{2}\Delta\alpha^t\,,
\end{equation}
where $\Delta\alpha^s=\alpha_{5/2}^s-\alpha_{7/2}^s$ and $\Delta\alpha^t=\frac{5}{7}\alpha_{7/2,4}^t-\frac{4}{5}\alpha_{5/2,3}^t$. By
definition, at the magic wavelength $\lambda_m$ the differential polarizability vanishes: $\Delta\alpha(\lambda_m)=0$.

The frequency shift of the clock transition due to the optical lattice can be caused by (i) accuracy of the magic wavelength determination and (ii) angular dependency of the tensor part of the differential polarizability.
The accuracy of the magic wavelength determination is related to the slope of $\Delta\alpha(\lambda)$ in the vicinity of $\lambda_m$, which is $-0.055(7)$\,a.u/nm for $\lambda_m\approx813$\,nm in Tm, as it is shown in the inset in Fig.\,\ref{img:mwl}.
In more practical units, this corresponds to the clock transition frequency shift of $U\times\Delta f\times 0.30(4) $\,mHz for lattice frequency detuning $\Delta f$\,[GHz] from $\lambda_m$ and  lattice depth $U$ in units of the recoil energy. 
This value is more than one order of magnitude smaller compared to the corresponding sensitivity of  Sr and Yb lattice clocks~\cite{Barber2008,brown2017hyperpolarizability}.

For $\Theta \ll 1$, the differential tensor polarizability $\Delta\alpha^t$ influences the clock transition frequency as:
\begin{equation}
\label{eq:tensor_shift}
h\Delta\nu  \approx -3/2\Delta\alpha^t \frac{E^2}{2}\Theta^2\,.
\end{equation}

To find $\Delta\alpha^t$, we measured dependency of the clock transition frequency shift $\Delta\nu$ on a small magnetic field $B_y$ at the constant  bias fields $B_x=225\,$mG and  $B_z=0$ ($\Theta\approx B_y/B_x$)  shown in  Fig.\,\ref{img:zeeman}(b).
From the corresponding parabolic coefficient of  56(11)\,mHz/mG${}^2$,  we get the differential tensor polarizability of $\Delta\alpha^t = 0.9(2)$\,a.u. at $\lambda_m$.
The uncertainty comes from the absolute calibration of magnetic field and power calibration~(see~Methods).
Both lattice frequency shifts mentioned in the previous paragraph can be readily reduced to mHz level by stabilizing the lattice wavelength with 0.1\,GHz accuracy and by maintaining  $|\Theta| < 10^{-3}$.

\begin{figure} [t!]
  \includegraphics[width=.8\linewidth]{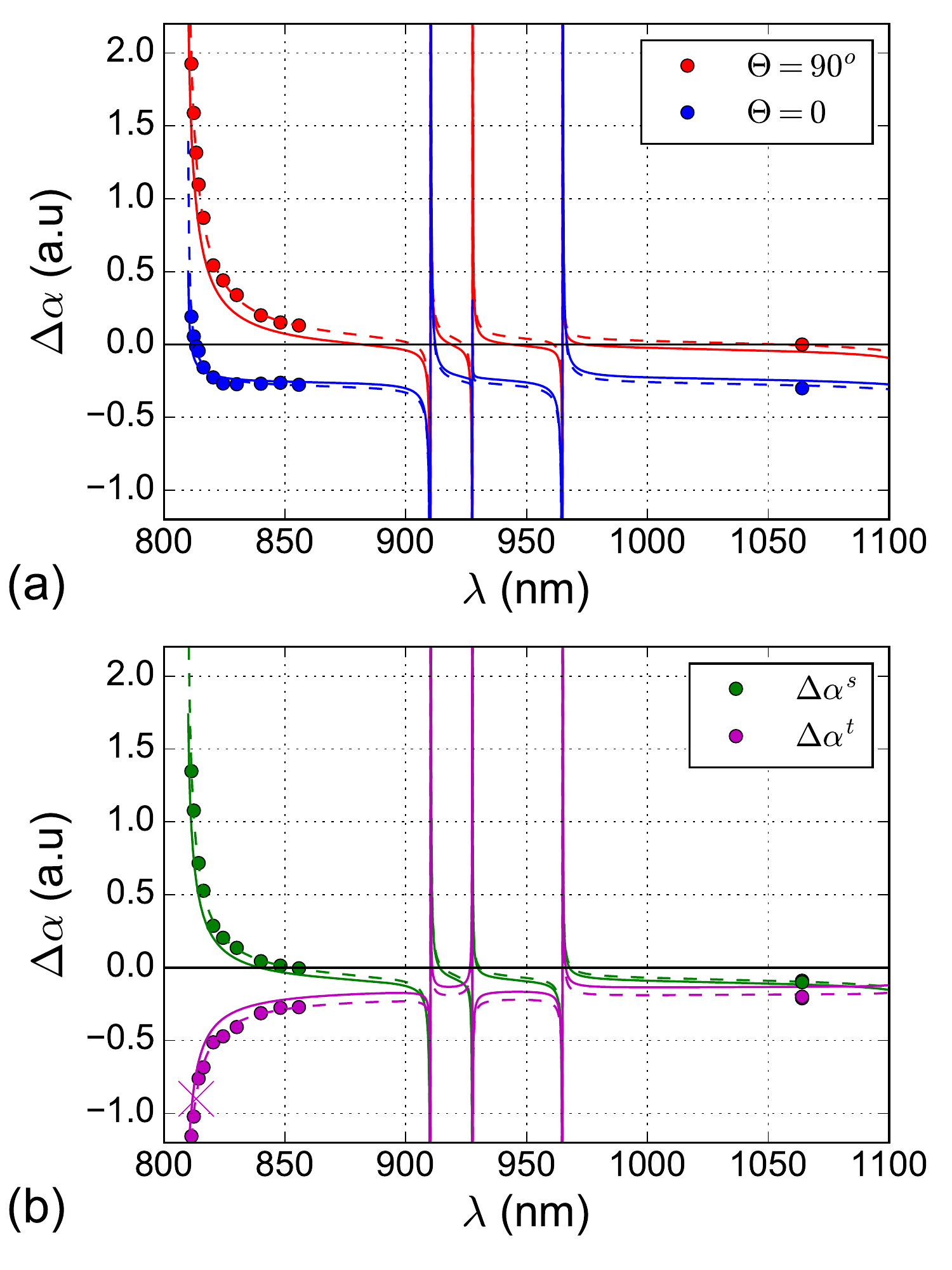}
\caption{Differential polarizabilities spectra.
a) The differential dynamic polarizability $\Delta\alpha(\lambda)$ for $\Theta=\pi/2$ (red dots) and  $\Theta=0$ (blue dots).
b)~Corresponding scalar $\Delta\alpha^s(\lambda)$ (green dots)  and tensor $\Delta\alpha^t(\lambda)$ (magenta dots) parts.
Magenta cross is $\Delta\alpha^t(\lambda_m)$ determined from measurements in Fig.\,\ref{img:zeeman}(b).
Solid and dotted curves are calculations based on Model 1 and Model 2, respectively~(see~Methods).
}
\label{img:pol_all}
\end{figure}

The frequency of $\ket{F=4,m_F=0}\rightarrow\ket{F=3,m_F=0}$ clock transition possesses a quadratic sensitivity to a dc magnetic field $B$ with a coefficient  $\beta=-257.2$\,Hz/G${}^2$~\cite{Sukachev2016}. 
To provide uncertainty of the transition frequency below 1\,mHz, it would be necessary to stabilize magnetic field at the level of $20\,\mu$G at the bias field of $B=100$\,mG.
Note, that the quadratic Zeeman shift in Tm can be fully canceled by measuring an averaged frequency of two clock transitions  $\ket{F=4,m_F=0}\rightarrow\ket{F=3,m_F=0}$ and $\ket{F=3,m_F=0}\rightarrow\ket{F=2,m_F=0}$ [Fig.\,\ref{img:setup}(b)] possessing the quadratic Zeeman coefficients of the opposite signs.
To implement this approach, one should provide  the magic-wavelength condition for both transitions. 
This can be done by choosing $\Theta=\arccos{(1/\sqrt{3})}$ to cancel the tensor part in~Eq.\,(\ref{eq:pol3}) and tuning the lattice wavelength approximately to 850\,nm [Fig.\,\ref{img:pol_all}(b)].
At this wavelength the differential scalar polarizability vanishes for both transitions ($\Delta\alpha^s=0$  since  $\alpha^s_{J,F}=\alpha^s_{J}$).

{\bf Static differential polarizability and the BBR shift.} The BBR shift of the clock transition frequency can be accurately calculated from the static differential scalar polarizability 
$\Delta\alpha^s_\textup{DC}=\Delta\alpha^s(\lambda\rightarrow\infty)$ from a theoretical model based on the measured polarizability spectrum at the wavelengths of 810--860\,nm and at 1064\,nm. 

Measurements in the spectral region of 810--860\,nm  were done by scanning the wavelength of the Ti:sapphire laser at two polarizations corresponding to
$\Theta=0$ ($\vec\epsilon\parallel\vec x $) and $\Theta=\pi/2$ ($\vec\epsilon\parallel\vec y $) as shown in~Fig.\,\ref{img:pol_all}(a).
The corresponding scalar $\Delta\alpha^s(\lambda)$ and tensor $\Delta\alpha^t(\lambda)$ differential polarizabilities calculated from Eq.\,(\ref{eq:pol3}) are shown in Fig.\,\ref{img:pol_all}(b).

To measure dynamic polarizability at 1064\,nm, we used a slightly different procedure: Tm atoms were trapped in the optical lattice at
$\lambda_m$ for which the differential polarizability vanishes, and the atomic cloud was illuminated along $y$-axis by a focused beam
of  a linearly-polarized single-frequency 1064\,nm fiber laser with the optical power up to 10\,W.
Corresponding results for $\Theta=0$ ($\vec\epsilon\parallel\vec x $) and $\Theta=\pi/2$ ($\vec\epsilon\parallel\vec z $) are also
shown in Fig.\,\ref{img:pol_all}.

To compare with experimental data and to deduce $\Delta\alpha^s_\textup{DC}$, we use Model\,2 (see~Methods), which differs from Model\,1 by introducing four adjustable parameters: probabilities of the 806.7\,nm and 809.5\,nm transitions and two offsets for differential scalar and tensor polarizabilities.
Corresponding fits based on Model\,2 are shown as dashed lines in  Fig.\,\ref{img:pol_all}, while calculations with no free parameters (Model\,1) are shown as solid lines.
From Model~2 we obtain
\begin{equation}\label{eq:polresult}
\Delta\alpha^s_\textup{DC}=-0.047(18)\,\textrm{a.u.}
\end{equation}
The differential static scalar polarizability from Model\,1 is $-0.062$\,a.u., which differs by only 1 standard deviation from~(\ref{eq:polresult}).
Calculations of the BBR frequency shift can be readily done using the value of $\Delta\alpha^s_\textup{DC}$~\cite{Sukachev2016}. 
The differential scalar polarizability in the spectral region around $10\,\mu$m (the maximum of the BBR spectrum at the room temperature) differs by less than $10^{-3}$\,a.u. from $\Delta\alpha_\textrm{DC}^s$. 
Note, that there are no  resonance transitions from the clock levels for $\lambda>1.5$\,$\mu$m.
For our  clock transition at $1.14\,\mu$m the room temperature  BBR shift is  $-0.45(18)$\,mHz. 
It is a few orders of magnitude smaller than for other neutral atoms and is comparable to the best ions species as shown in Table\,\ref{bbr_table}.
This result quantitatively confirms the idea of strong shielding of inner-shell transitions in lanthanides from external electric fields.

The clock transition frequency shifts due to a magnetic component of the BBR and electric field gradient of the optical lattice are less than $10^{-4}$\,Hz and can be neglected (see Methods).

\begin{table}[t!]
\caption{The fractional BBR shift at 300\,K for the clock transition frequencies
in thulium and some other neutral atoms and ions.}
\begin{ruledtabular}
\begin{tabular}{cccc}
 &Element & $\Delta\nu^{BBR}/\nu, 10^{-17}$ \\
\hline
 &Tm (this work)  & $-0.2$\\
 &Sr \footnote[1]{ref.\,\cite{Ludlow2015}}  & $-550$\\
  &Yb \footnotemark[1] & $-270$ \\
   &Hg \footnote[2]{ref.\,\cite{Bilicki2016}}  & $-16$ \\
  &Yb$^+$ (E3)\footnotemark[1]  & $-11$\\
  &Al$^+$ \footnotemark[1]   & $-0.4$ \\
  &Lu$^+$ \footnote[3]{transition ${}^1S_0 - {}^3D_1$, ref.\,\cite{Arnold2018}}  &  $-0.14$ \\
\end{tabular}

\end{ruledtabular}\label{bbr_table}
\end{table}

\subsection*{Discussion}

Specific shielding of the inner-shell magnetic-dipole clock transition in atomic thulium at 1.14\,$\mu$m by outer 5$s^2$ and 6$s^2$  electronic shells results in a very low sensitivity of its frequency to external electric fields. 
The differential static scalar polarizability of the two  $J=7/2$ and $J=5/2$ clock levels is only $-0.047(18)$\,atomic units which corresponds to the fractional BBR frequency shift of the transition of $2\times 10^{-18}$ at the room temperature. It is  by three orders of magnitude less compared to the prominent clock transitions in Sr and Yb (see Table\,\ref{bbr_table}). 
Taking into account that all major frequency shifts (the Zeeman shift, lattice shifts, collisional shifts) can be controlled at the low $10^{-17}$ level, these features
make thulium a promising candidate for a transportable room-temperature optical atomic clock due to soft constrains on the ambient temperature stability. 
It combines advantages of unprecedented frequency stability of optical lattice clocks on neutral atoms  and low sensitivity to BBR of ion optical clocks.
Moreover, precision spectroscopy in Tm opens possibilities for sensitive tests of Lorentz invariance \cite{Shaniv2018} and for search of the fine structure constant variation \cite{kolachevsky2008high}.

Optical clocks based on a $f$-$f$ transition in some other lanthanides with spinless nuclei could be even more attractive featuring the low sensitivity to magnetic fields due to the absence of the hyperfine structure and small BBR shift.
For example, the fine-structure clock transition at the telecom-wavelength of 1.44\,$\mu$m in laser-cooled erbium atoms (e.g. $^{166}$Er)~\cite{McClelland2006,Kozlov2013} can be particularly interesting for optical frequency dissemination
over fiber networks~\cite{Riehle2017}.

\subsection*{Methods}
{\bf Enhancement cavity}. Optical lattice is formed inside a $\Gamma$-shaped enhancement cavity, as shown in Fig.\,\ref{img:setup}(a).
The reflectivity  of the curved (the radius is $r=-250$\,mm) incoupler mirror M1 equals 87\% and matches losses introduced by the
vacuum chamber viewports. Outcouplers M2 or M2$^\prime$ are  identical flat mirrors  with
the reflectivity of $R>99\%$.
For locking to the laser frequency, the  cavity  mirror reflecting the beam at $45^\circ$ is mounted on a piezo actuator.
The intra-cavity polarization is defined by a broadband polarization beam splitter; depending on the polarization, either M2 or
M2$^\prime$ outcoupler
mirror is used. The  intra-cavity lens has the focal length of $f=400$\,mm. 

Depending on experimental geometry ($\Theta=0$ or $\Theta=\pi/2$), we couple corresponding linearly polarized radiation from the
Ti:sapphire
laser through the incoupler mirror~M1.
Intra-cavity polarization filtering by PBS defines the polarization angle and  significantly improves polarization purity of the
optical lattice. The angle between the laser field polarization and the bias magnetic field is adjusted with the accuracy of  better than $1^\circ$.

{\bf Measurement of differential dynamic polarizabilities.} Differential polarizability $\Delta\alpha$ of the clock levels is determined from the frequency shift of the corresponding transition
$\Delta\nu$, circulating power $P$, and TEM$_{00}$ cavity mode  radius $w$ at the atomic cloud position as

\begin{equation}\label{eq4}
\Delta\alpha =-\frac{hc w^2}{16 a_0^3}\frac{\Delta\nu}{P}\,.
\end{equation}

\begin{figure}[h!]
	\includegraphics[width=.8\linewidth]{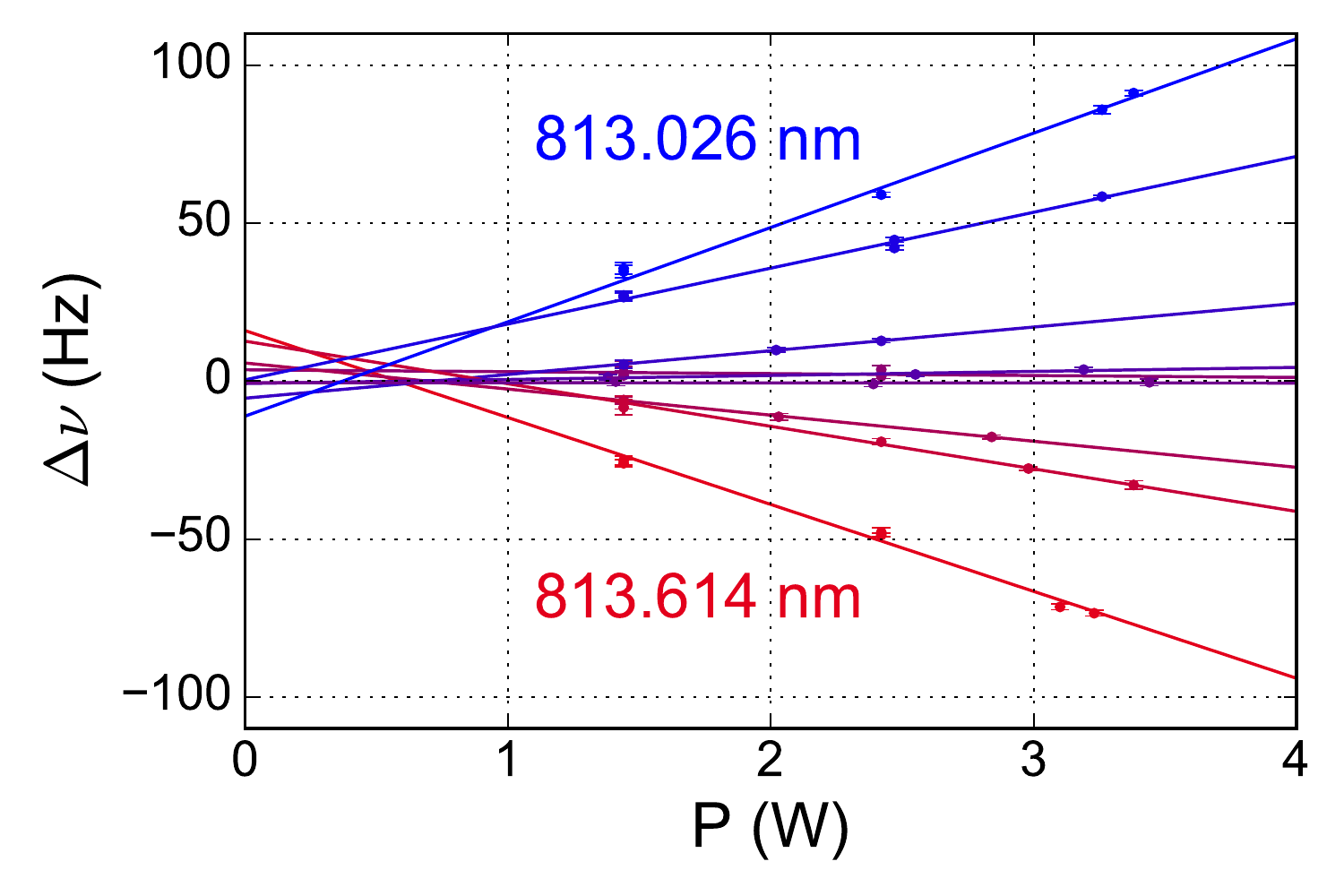}
\caption{
 Clock transition frequency shift $\Delta\nu$ as a function of optical lattice power $P$ in the vicinity of the magic
wavelength $\lambda_m=813.32$\,nm. Experimental data are fitted by linear functions.}
\label{img:s2}
\end{figure}

The dependency $\Delta\nu(P)$ was obtained for different wavelengths in the spectral range  810--860\,nm for the intra-cavity circulating power varying from 1\,W to 4\,W, as it is shown in Fig.\,\ref{img:s2}.
Frequency shifts $\Delta\nu$  were measured relative to the laser frequency, which is stabilized to an ultra-stable ULE cavity with
linear  drift compensation.
The slope coefficients of corresponding linear fits were substituted to Eq.\,(\ref{eq4}) to deduce $\Delta\alpha$ presented in Figs.\,\ref{img:mwl},\,\ref{img:pol_all}.

\begin{figure}[h!]
    \includegraphics[width=.8\linewidth]{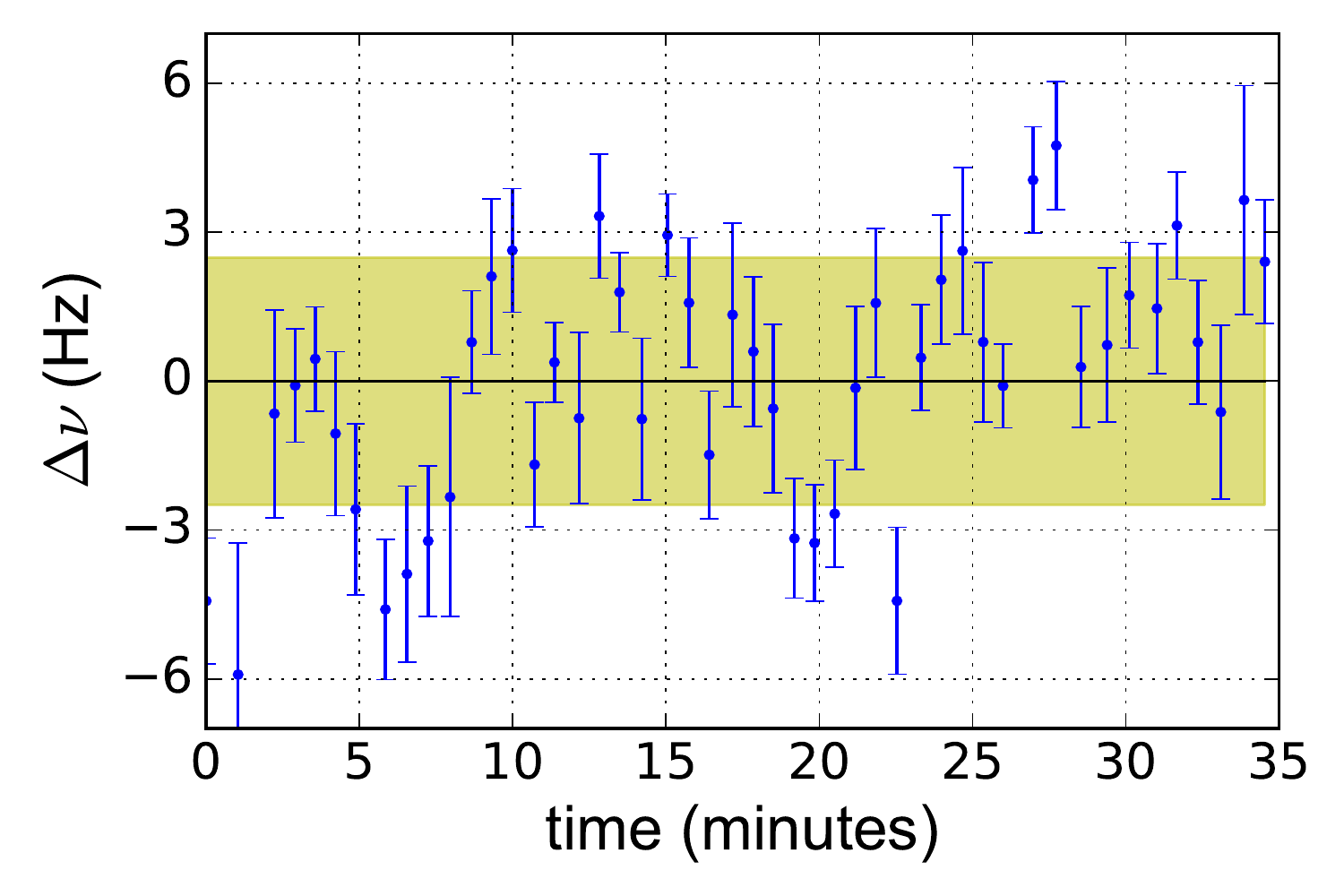}
\caption{Relative frequency of Tm clock transition and ULE cavity mode with linear drift compensation. Each data point and
corresponding uncertainty comes from the fit of the clock transition spectrum. The shaded region corresponds to 1~standard deviation of the data set.}
\label{img:s3}
\end{figure}

Uncertainty of the frequency shift $\Delta\nu$ comes from the residual instability of the reference cavity on time intervals of
1000\,s.
To estimate it, we measured clock transition frequency relative to the clock laser frequency at the magic wavelength when the
perturbations from the lattice are minimal. Results are shown in  Fig.\,\ref{img:s3}. The standard deviation equals to 2.6\,Hz
contributing $0.003$\,a.u. to the error budget of  $\Delta\alpha$. For the lattice wavelength  detuned from $\lambda_m$ contribution
of the laser frequency instability  is negligible.

The intra-cavity power $P$ was determined by measuring power leaking through the cavity outcoupler  M2 (or M2$^\prime$) using
calibrated photodiodes. For each photodiode we measured the power-to-voltage transfer function $P(U) = \kappa  U$,  where $U$ is the
voltage reading from the photodiode, and $\kappa$ is the coefficient measured using absolutely calibrated Thorlabs S121C power meter.
To determine $\kappa$, we unlocked the cavity, slightly tilted the outcoupler, blocked the reflected beam to prevent feasible
reflections from the incoupler, and measured the power before the outcoupler  and corresponding voltage reading of the photodiode.
The
linearity of the photodiode response was checked separately and turned out to be better than $3\%$ in the working region. The
photodiode calibration was done in the whole spectral range of 810--860\,nm to take into account the spectral response of the outcoupler
and the photodiode.
Although the specified uncertainty of the power sensor Thorlabs S121C equals 3\%, we ascribe the net uncertainty
of power measurement of 10\% from comparison of  readings from three different absolutely calibrated sensors.

The beam radius $w$ at the atomic cloud position is deduced from the cavity geometry: distances from the vacuum chamber center (and target
atomic cloud position) to M1, L and M2 (or M2$^\prime$) are 244\,mm, 384\,mm and 500\,mm, respectively, giving the beam radius of
$w=126\,\mu$m at the position of atomic cloud.
The uncertainty of $w$ comes from the position uncertainties of cavity elements and of the atomic cloud with respect to the chamber
center, as well as from uncertainty of 1\,mm of the lens focal length.
We conservatively evaluate position uncertainties of  M1, L, and M2 (or PBS and M2$^\prime$) as 1\,mm, and the possible axial
displacement of the atomic cloud of 2\,mm.
The partial contributions to the beam radius uncertainty are $1.6\,\mu$m (the incoupler), $1.2\,\mu$m (the lens), $0.03\,\mu$m (the
outcoupler), $0.3\,\mu$m (the cloud) and $1.5\,\mu$m (focal length).
Adding up in quadratures,  the total  uncertainty of the beam radius $w_0$ equals $2.5\,\mu$m.
The result is independently confirmed by measuring frequency intervals  between cavity transversal modes.

The finite temperature of the atomic cloud reduces  averaged light intensity on the atomic cloud $I_\textrm{av}$  in respect to the
on-axis antinode lattice intensity  $I_0=8P/\pi w^2$.
 Assuming  Boltzman distribution of the atoms with the temperature $T$ in the trap of depth $U_0$ one can calculate the parameter
 $\eta$ connecting the averaged and maximum intensity   $I_\textrm{av} = \eta I_0$. The $\eta$ parameter is calculated from 
\begin{equation}
\eta = \frac{\int\limits_{0}^{U_0} e^{-E/kT} \left(\frac{1}{2 r_0} \int\limits_{-r_0}^{r_0}
e^{-2r^2}dr\right)\,dE}{\int\limits_{0}^{U_0} e^{-E/kT}dE}\,,
\end{equation}
where $r_0  = ({-\ln(1-E/U_0)/2})^{-1/2}$ is the classical turning point.
The parameter $\eta$ equals $0.90(5)$ for $kT=0.3\,U_0$ (which corresponds  to average experimental conditions).
One standard deviation $\sigma=0.05$ of $\eta=0.9$ corresponds to $kT$ range from $0.14\,U_0$ to $0.6\,U_0$, while  $2\,\sigma$ already covers
($0$\,-\,$3)\,U_0$ range.

Summarizing,  we evaluate the uncertainty of measured differential polarizability  $\Delta\alpha$ to be 13\% with the dominating
contribution from power calibration.
In the vicinity of $\lambda_m$, the uncertainty is slightly higher by 0.003\,a.u. because  of the reference clock laser
frequency instability.

To measure differential polarizability at 1064\,nm, atoms were trapped in the optical lattice at the magic wavelength and irradiated
by focused slightly elliptic 1064\,nm laser beam with the waists of $w_x=320(20)\,\mu$m and $w_y=280(20)\,\mu$m in  $x$ and $y$
directions, respectively. To adjust the 1064\,nm laser beam center to the atomic cloud, we maximized intensity of the beam  on the
atomic cloud by monitoring the frequency shift of the clock transition. To increase sensitivity, corresponding measurements were done
by tuning the clock laser to the slope of 1.14\,$\mu$m transition.

For the  measurement session of $\Delta\alpha$ at $\Theta=0$,  we performed three adjustments of 1064\,nm laser beam. The
reproducibility corresponds to the frequency shift of 5\,Hz at the maximal frequency shift of 25\,Hz. The resulting linear
coefficient
is evaluated as $\Delta\nu/P=-3.1(6)$\,Hz/W including this uncertainty.
In $\Theta=\pi/2$ configuration, we did not observe any significant  effect from adjustment and the coefficient equals
$0.04(24)$\,Hz/W.


{\bf Theoretical  analysis.} Theoretical approach to calculate  polarizabilities of Tm atomic levels  is described in our previous works~\cite{Sukachev2016,golovizin2017methods}. Calculations are based on the time-dependent second order perturbation theory with summation over known discrete transitions from the levels of interest. For calculations, we  used transitions wavelengths and probabilities obtained
with the {\small COWAN} package \cite{cowan1981} with some exceptions: for transitions with $\lambda>800$\,nm,  experimental wavelengths from ref.\,\cite{NIST_ASD} are used. This approach allows to increase  accuracy
of the magic wavelength prediction in the corresponding spectral region of $\lambda>800$\,nm. According to calculations, the magic
wavelength  was expected at 811.2\,nm which motivated our experimental studies in the spectral region 810--815\,nm (see
Fig.\,\ref{img:mwl}). We refer to this model as ``Model\,1'' and use for comparison with experimental results of this work as
shown  in Figs.\,\ref{img:mwl},\ref{img:pol_all}.

Deviation of the  experimental data from Model\,1 can be explained by two main factors. 
First, in this model we did not take into account transitions to the continuum. 
Together with  uncertainties of  {\small COWAN} calculations of transition amplitudes, it can result in a small offset of the infrared differential polarizability spectrum. 
Note, that although transitions to the continuum spectrum may significantly contribute to polarizabilities of the individual levels (up to 10\%), for the differential polarizability of the $f$-$f$ transition at 1.14\,$\mu$m in Tm contribution of the continuum is mostly canceled \cite{Sukachev2016}.

\begin{table}[h]
\caption{Uncertainty budget for the differential scalar polarizability $\Delta\alpha^s_\textup{DC}$.}
\begin{ruledtabular}
\begin{tabular}{l|c}
Source & Uncertainty, a.u.\\
\hline
 Experimental results for  810--850\,nm  &  $0.013$ \\
 Experimental result for  $1064$\,nm  &  $0.006$ \\
 Angle $\Theta$ &  $0.002$\\
  Transition probabilities for $\lambda>900$\,nm   &  $0.01$ \\
   \hline
  \bf{Total}   &    $\bf{0.018}$ \\

  \hline
    For the reference: & \\
    Difference of Models 1 and 2&   $0.015$\\
\end{tabular}
\end{ruledtabular}\label{table2}
\end{table}

To fit experimental data we use Model\,2 which differs from Model\,1 by introducing four fit parameters. As parameters we use the probabilities of the 806.7\,nm and 809.5\,nm transitions which mostly effect polarizability spectrum in 810--860\,nm region and two offsets for the scalar and  tensor polarizabilities. 
After fitting the experimental data (see Fig.\,\ref{img:pol_all}) by Model\,2, the  probability of the 806.7\,nm transition is changed from $3473\,\text{s}^{-1}$ to $4208(298)\,\text{s}^{-1}$, the probability of the 809.5\,nm transition is changed from 149\,s$^{-1}$ to 357(109)\,s$^{-1}$, the fitted offsets for the differential scalar and tensor polarizabilities equal $0.012(6)$\,a.u. and $-0.028(12)$\,a.u. with reduced $\chi_s^2$ for the fits of 1.35 and 2.9, respectively.

Transitions from the upper Tm clock level $J=5/2$  in the spectral range $\lambda>900$\,nm  are weak and their probabilities are experimentally not measured.  
To calculate probabilities we used  {\small COWAN} package. To estimate the impact of insufficient knowledge of transition probabilities on differential scalar polarizability $\Delta\alpha^s(\lambda)$, we assume possible variation of each transition  probability by a factor of 2. 
After extrapolation of the fitted Model\,2 to $\lambda\rightarrow\infty$ we get the static differential polarizability $\Delta\alpha^s_\textup{DC}=-0.047^{+0.01}_{-0.005}$\,a.u., where uncertainty comes from variation of $\lambda>900$\,nm transition probabilities.

We summarize all sources of uncertainties which contribute to the error of the static differential polarizability $\Delta\alpha^s_\textup{DC}$ in Table\,\ref{table2}.
As discussed above, the experimental uncertainty for 810--860\,nm range is 13\% contributing $0.013$\,a.u. to the $\Delta\alpha^s_\textup{DC}$, while the measurement at 1064\,nm is less accurate (20\%) due to the laser beam adjustment and results in  $0.006$\,a.u. variation of Model\,2 extrapolation. The uncertainty of the angle $\Theta$ adjustment  contributes $0.002$\,a.u.
The uncertainty coming from the poorly known transition probabilities from $J=5/2$ clock level in $\lambda>900$\,nm range contributes $0.01$\,a.u.
Using extrapolation of Model\,2 and adding all uncertainties we come to the  final result of $\Delta\alpha^s_\textup{DC}=-0.047(18)$\,a.u.
Note, that this is fully consistent with the extrapolated value of -0.062\,a.u. obtained from Model\,1 and given for the
reference in Table\,\ref{table2}.


{\bf BBR magnetic field.} To estimate the clock  transition frequency shift due to the magnetic component of BBR, we follow the analysis given in the work~\cite{gan2018oscillating}. 
The corresponding frequency shift of one of the  clock levels coupled to another atomic level with magnetic-dipole transition at frequency $\omega_0$ can be found by integrating over the full BBR spectrum as
\begin{equation}
\begin{split}
\Delta\nu^B_{bbr}(T) &= -\frac{\omega_0 }{2\pi}\frac{\mu_B^2}{2 \hbar \pi^2 c^5 \epsilon_0}\int_0^\infty \frac{1}{\omega_0^2-\omega^2}\frac{\omega^3}{e^{\hbar \omega/kT} -1 }d\omega \\
& = -\frac{\omega_0 }{2\pi}\frac{\gamma}{2} \left(\frac{T}{T_0}\right)^2 f(y),
\end{split}
\end{equation}
where $\epsilon_0$ is the vacuum permittivity, $T_0=300$\,K, $y=\hbar\omega_0/k_B T$. Here
\begin{gather}
\gamma = \frac{\mu_B^2}{\hbar^2}\frac{\hbar}{6 c^5 \epsilon_0} \left(\frac{k_B T_0}{\hbar}\right)^2 \approx 9.78\times10^{-18}, \\
f(y) = \frac{6}{\pi^2}\int_0^\infty\frac{1}{y^2-x^2}\frac{x^3 dx}{e^x-1}.
\end{gather}

The hyperfine transition frequency $\omega_0$ of the $J=7/2$ ground level in Tm equals $2\pi\times1496$\,MHz, while for the $J=5/2$ clock level it equals $2\pi\times2115$\,MHz. 
For these values of $\omega_0$ $y \ll 1$, $f(y)\approx -1$, and the shift is on the order of $10^{-8}$\,Hz.
To estimate the contribution from the optical transitions, we evaluate the shift from the lowest frequency magnetic-dipole transition at $\omega_0 \approx 2\pi\times263$\,THz, which is the clock transition itself: $y=42$, $f(y)=2.3\times10^{-3}$ and $\Delta\nu^B_{bbr}(T_0)=-3\times10^{-6}$\,Hz for the ground level (for the upper clock level the corresponding shift is $+3\times10^{-6}$\,Hz).
Hence, we estimate total shift of the clock transition from the magnetic component of BBR to be less than $10^{-4}$\,Hz.

{\bf Electric quadrupole shift.} Opposite to neutral Sr, Yb, and Hg atoms, Tm clock levels posses a non-zero electric quadrupole moment on the order of 1\,a.u.~\cite{Sukachev2016} and are coupled to an electric field gradient. 
Since the electric field gradient in an optical lattice oscillates at the optical frequency, the corresponding time-averaged frequency shift of the clock transition is zero. 

\subsection*{References}

\begin{thebibliography}{10}

\bibitem{Chen2017}
{Chen, J. et al.}
\newblock {Sympathetic ground state cooling and time-dilation shifts in an
  ${}^{27}$Al${}^+$ optical clock}.
\newblock {\em Phys. Rev. Lett.} {\bf 118,} 053002 (2017).

\bibitem{Ushijima2015}
{Ushijima, I., Takamoto, M., Das, M., Ohkubo, T. \& Katori, H.}
\newblock {Cryogenic optical lattice clocks}.
\newblock {\em Nature Photonics} {\bf 9,} 185--189 (2015).

\bibitem{Marti2018}
{Marti, G.~E. et al.}
\newblock Imaging optical frequencies with $100\text{ }\text{
  }\ensuremath{\mu}\mathrm{Hz}$ precision and $1.1\text{ }\text{
  }\ensuremath{\mu}\mathrm{m}$ resolution.
\newblock {\em Phys. Rev. Lett.} {\bf 120,} 103201 (2018).

\bibitem{Delva2017}
{Delva, P. et al.}
\newblock Test of special relativity using a fiber network of optical clocks.
\newblock {\em Phys. Rev. Lett.} {\bf 118,} 221102 (2017).

\bibitem{Shaniv2018}
{Shaniv, R. et al.}
\newblock New methods for testing lorentz invariance with atomic systems.
\newblock {\em Phys. Rev. Lett.} {\bf 120,} 103202 (2018).

\bibitem{roberts2017search}
{Roberts, B.~M. et al.}
\newblock Search for domain wall dark matter with atomic clocks on board global positioning system satellites.
\newblock {\em Nature Communications} {\bf 8,} 1195 (2017).

\bibitem{Wciso2018}
{Wcis{\l}o, P. et al.}
\newblock {First observation with global network of optical atomic clocks aimed for a dark matter detection}.
\newblock {\em arXiv preprint arXiv:1806.04762} (2018).

\bibitem{Rosenband2008}
{Rosenband, T. et al.}
\newblock Frequency ratio of Al$^+$ and Hg$^+$ single-ion optical clocks; metrology at the 17th decimal place.
\newblock {\em Science} {\bf 319,} 1808--1812 (2008).

\bibitem{Borkowski2018}
{Borkowski, M.}
\newblock Optical lattice clocks with weakly bound molecules.
\newblock {\em Phys. Rev. Lett.} {\bf 120,} 083202 (2018).

\bibitem{Norcia259}
{Norcia, M.~A., et al.}
\newblock Cavity-mediated collective spin-exchange interactions in a strontium
  superradiant laser.
\newblock {\em Science} {\bf 361,} 259--262 (2018).

\bibitem{Ludlow2015}
{Ludlow, A., Boyd, M., Ye, J., Peik, E. \& Schmidt, P.}
\newblock {Optical atomic clocks}.
\newblock {\em Reviews of Modern Physics} {\bf 87,} 637 (2015).

\bibitem{Poli2014}
{Poli, N., Oates, C.~W., Gill, P. \& Tino, G.~M.}
\newblock {Optical atomic clocks}.
\newblock {\em arXiv preprint arXiv:1401.2378\/} (2014).

\bibitem{safronova2018search}
{Safronova, M.~S. et al.}
\newblock Search for new physics with atoms and molecules.
\newblock {\em Rev. Mod. Phys.} {\bf 90,} 025008 (2018).

\bibitem{Koller2017}
{Koller, S.~B. et al.}
\newblock Transportable optical lattice clock with
  $7\ifmmode\times\else\texttimes\fi{}{10}^{\ensuremath{-}17}$ uncertainty.
\newblock {\em Phys. Rev. Lett.} {\bf 118,} 073601 (2017).

\bibitem{Grotti2018}
{Grotti, J. et al.}
\newblock {Geodesy and metrology with a transportable optical clock}.
\newblock {\em Nature Physics} {\bf 14,} 437--441 (2018).

\bibitem{takano2016geopotential}
{Takano, T. et al.}
\newblock Geopotential measurements with synchronously linked optical lattice clocks.
\newblock {\em Nature Photonics} {\bf 10,} 662--666 (2016).

\bibitem{Nicholson2015}
{Nicholson, T. et al.}
\newblock {Systematic evaluation of an atomic clock at $2\times10^{-18}$ total uncertainty.}
\newblock {\em Nature communications} {\bf 6,} 6896 (2015).

\bibitem{schioppo2017ultrastable}
{Schioppo, M. et al.}
\newblock Ultrastable optical clock with two cold-atom ensembles.
\newblock {\em Nature Photonics} {\bf 11,} 48--52 (2017).

\bibitem{Huntemann2016}
{Huntemann, N., Sanner, C., Lipphardt, B., Tamm, C. \& Peik, E.}
\newblock {Single-ion atomic clock with $3\times10^{-18}$ systematic uncertainty}.
\newblock {\em Phys. Rev. Lett.} {\bf 116,} 063001 (2016).

\bibitem{Chou2010}
{Chou, C., Hume, D., Koelemeij, J., Wineland, D. \& Rosenband, T.}
\newblock {Frequency comparison of two high-accuracy Al$^+$ optical clocks}.
\newblock {\em Phys. Rev. Lett.} {\bf 104,} 070802 (2010).

\bibitem{Arnold2018}
\newblock {Arnold, K.~J., Kaewuam, R., Roy, A., Tan, T.~R., \& Barrett, M.~D.}
\newblock {Blackbody radiation shift assessment for a lutetium ion clock}.
\newblock {\em Nature Communications} {\bf 9,} 1650 (2018).

\bibitem{nauta2017towards}
{Nauta, J. et al.}
\newblock Towards precision measurements on highly charged ions using a high harmonic generation frequency comb.
\newblock {\em Nuclear Instruments and Methods in Physics Research Section B: Beam Interactions with Materials and Atoms} {\bf 408,} 285--288 (2017).

\bibitem{Yu2018}
{Yu, Y.-m. \& Sahoo, B.~K.}
\newblock Selected highly charged ions as prospective candidates for optical
  clocks with quality factors larger than ${10}^{15}$.
\newblock {\em Phys. Rev. A} {\bf 97,} 041403 (2018).

\bibitem{Kozlov2018}
{Kozlov, M.~G., Safronova, M.~S., L{\'{o}}pez-Urrutia, J. R.~C. \&
  Schmidt, P.~O.}
\newblock {Highly charged ions: optical clocks and applications in fundamental
  physics}.
\newblock {\em arXiv preprint arXiv:1803.06532} (2018).

\bibitem{Wense2017}
{Wense, L. V.~D. et al.}
\newblock {A laser excitation scheme for ${}^{229\textrm{m}}$Th}.
\newblock {\em Phys. Rev. Lett.} {\bf 119,} 132503 (2017).

\bibitem{thielking2018laser}
{Thielking, J. et al.}
\newblock Laser spectroscopic characterization of the nuclear-clock isomer ${}^{229\textrm{m}}$Th.
\newblock {\em Nature} {\bf 556,} 321--325 (2018).

\bibitem{aleksandrov1984eb}
{Aleksandrov, E.~B., Vedenin, V.~D. \& Kulyasov, V.~N.}
\newblock Broadening and shift of thulium resonance lines by helium.
\newblock {\em Opt. Spectrosc.} {\bf 56,} 365--368 (1984).

\bibitem{Hancox2004}
{Hancox, C.~I., Doret, S.~C., Hummon, M.~T., Luo, L. \& Doyle, J.~M.}
\newblock {Magnetic trapping of rare-earth atoms at millikelvin temperatures}.
\newblock {\em Nature} {\bf 431,} 281--284 (2004).

\bibitem{Hancox2005}
{Hancox, C.~I., Doret, S.~C., Hummon, M.~T., Krems, R.~V. \&
  Doyle, J.~M.}
\newblock {Suppression of angular momentum transfer in cold collisions of
  transition metal atoms in ground states with nonzero orbital angular
  momentum}.
\newblock {\em Phys. Rev. Lett.} {\bf 94,} 013201 (2005).

\bibitem{probst2013}
{Probst, S. et al.}
\newblock Anisotropic rare-earth spin ensemble strongly coupled to a
  superconducting resonator.
\newblock {\em Phys. Rev. Lett.} {\bf 110,} 157001 (2013).

\bibitem{Dibos2018}
{Dibos, A.~M., Raha, M., Phenicie, C.~M. \& Thompson, J.~D.}
\newblock {Atomic source of single photons in the telecom band}.
\newblock {\em Phys. Rev. Lett.} {\bf 120,} 243601 (2018).

\bibitem{Kozlov2013}
{Kozlov, A., Dzuba, V.~A. \& Flambaum, V.~V.}
\newblock Prospects of building optical atomic clocks using Er i or Er iii.
\newblock {\em Phys. Rev. A} {\bf 88,} 032509 (2013).

\bibitem{kozlov2014optical}
{Kozlov, A., Dzuba, V.~A. \& Flambaum, V.~V.}
\newblock Optical atomic clocks with suppressed blackbody-radiation shift.
\newblock {\em Phys. Rev. A} {\bf 90,} 042505 (2014).

\bibitem{Sukachev2016}
{Sukachev, D. et al.}
\newblock Inner-shell magnetic dipole transition in Tm atoms: A candidate for
  optical lattice clocks.
\newblock {\em Phys. Rev. A} {\bf 94,} 022512 (2016).

\bibitem{sukachev2014secondary}
{Sukachev, D.~D. et al.}
\newblock Secondary laser cooling and capturing of thulium atoms in traps.
\newblock {\em Quantum Electronics} {\bf 44,} 515--520 (2014).

\bibitem{Alnis2008}
{Alnis, J., Matveev, A., Kolachevsky, N., Udem, T. \& H\"ansch, T.~W.}
\newblock Subhertz linewidth diode lasers by stabilization to vibrationally and
  thermally compensated ultralow-expansion glass Fabry-P\'erot cavities.
\newblock {\em Phys. Rev. A} {\bf  77,} 053809 (2008).

\bibitem{Takamoto2003}
{Takamoto, M. \& Katori, H.}
\newblock {Spectroscopy of the ${}^1S_0-{}^3P_0$ clock transition of ${}^{87}$Sr in an  optical lattice}.
\newblock {\em Phys. Rev. Lett.} {\bf 91,} 223001 (2003).

\bibitem{cowan1981}
{Cowan, R.}
\newblock The theory of atomic structure and spectra.
\newblock (University of California Press, Berkeley, CA, 1981), and Cowan
  programs RCN, RCN2,and RCG.
  
\bibitem{Barber2008}
{Barber, Z.~W. et al.}
\newblock {Optical lattice induced light shifts in an Yb atomic clock.}
\newblock {\em Phys. Rev. Lett.} {\bf 100,} 103002 (2008).

\bibitem{brown2017hyperpolarizability}
{Brown, R.~C. et al.}
\newblock Hyperpolarizability and operational magic wavelength in an optical
  lattice clock.
\newblock {\em Phys. Rev. Lett.} {\bf 119,} 253001 (2017).

\bibitem{kolachevsky2008high}
{Kolachevsky, N.~N.}
\newblock High-precision laser spectroscopy of cold atoms and the search for
  the drift of the fine structure constant.
\newblock {\em Physics-Uspekhi} {\bf 51,} 1180--1190 (2008).

\bibitem{McClelland2006}
{McClelland, J.~J. \& Hanssen, J.~L.}
\newblock Laser cooling without repumping: A magneto-optical trap for erbium
  atoms.
\newblock {\em Phys. Rev. Lett.} {\bf 96,} 143005 (2006).

\bibitem{Riehle2017}
{Riehle, F.}
\newblock {Optical clock networks}.
\newblock {\em Nature Photonics} {\bf 11,} 25--31 (2017).

\bibitem{golovizin2017methods}
{Golovizin, A. et al.}
\newblock Methods for determining the polarisability of the fine structure
  levels in the ground state of the thulium atom.
\newblock {\em Quantum Electronics} {\bf 47,} 479--483 (2017).

\bibitem{NIST_ASD}
{Kramida, A., Ralchenko, Yu., Reader, J., and {NIST ASD Team}}.
\newblock {NIST Atomic Spectra Database (ver. 5.5.6), [Online]. Available:
  {\tt{https://physics.nist.gov/asd}} [2018, August 21]. National Institute of
  Standards and Technology, Gaithersburg, MD.}, 2018.
  
\bibitem{gan2018oscillating}
{Gan, H. et al.}
\newblock Oscillating magnetic field effects in high precision metrology.
\newblock {\em arXiv preprint arXiv:1807.00424\/} (2018).

\bibitem{Bilicki2016}
{Tyumenev, R. et al.}
\newblock Comparing a mercury optical lattice clock with microwave and optical frequency standards.
\newblock {\em New Journal of Physics} {\bf 18,} 113002 (2016).

\end{thebibliography}

\subsection*{Acknowledgment}
The work is supported by RFBR Grants No.\,18-02-00628
and No.\,16-29-11723. We are grateful to  S.\,Kanorski and V.\,Belyaev for invaluable technical support and to S.\,Fedorov for
assembling the frequency comb. We also thank M. Barrett for very useful discussions.

\subsection*{Author contributions}
A.G., E.F., and D.T. carried out all measurements and analyzed the data; A.G. and D.S. did theoretical calculations; A.G., D.S., and N.K. wrote the paper. K.K., V.S., and N.K. conceived and directed the project. All authors discussed the results and commented on the manuscript.

\subsection*{Additional information}
{\bf Competing interests:} The authors declare no competing interests.

\end{document}